\documentstyle[12pt]{article}

\begin{document}

\newcommand{\beq}{\begin{equation}}
\newcommand{\beqn} {\begin{eqnarray}}
\newcommand{\r}{{\vec r}}
\newcommand{\eeq}{\end{equation}}
\newcommand{\eeqn} {\end{eqnarray}}
\newcommand{\pa}{{p}^a}
\newcommand{\pbb}{{p}^b}
\newcommand{\amp}{{\cal A}}
\newcommand{\kta}{{\vec k_a}}
\newcommand{\ktb}{{\vec k_b}}
\newcommand{\ktc}{{\vec k_c}}
\newcommand{\q}{\Delta}
\newcommand{\qt}{\vec\Delta}
\newcommand{\nnn}{\noindent}
\renewcommand{\thefootnote}{\dag}
\newcommand{\numero}{McGill/93-35}
\newcommand{\numerodeux}{hep-ph/9406435}

\newcommand{\auteura}{J.R. Cudell\footnote[1]{cudell@hep.physics.mcgill.ca}
}
 \newcommand{\auteurb}{Department of Physics\\
McGill University\\ 3600 University Street\\
Montr\'eal, Qu\'ebec, H3A 2T8, Canada}

\newcommand{\titre}{ THE DONNACHIE-LANDSHOFF\\ POMERON VS.
QCD\footnote[2]{Talk presented at the $\rm V^{th}$ Blois Workshop on Elastic
and Diffractive Scattering, Brown University, Rhode Island, June 8--12,
1993.}}

\newcommand{\abstrait}{The Donnachie-Landshoff model of diffractive scattering,
which accounts for all
confirmed diffractive data, is briefly reviewed. The picture emerging
from this model is shown to directly contradict that coming from perturbative
leading-log {\it s} calculations. The inclusion of non-perturbative effects
brings the two pictures in closer agreement.}

\begin{titlepage}
\rightline{\numero}
\rightline{\numerodeux}\ \\
\vspace{.5in}
\begin{center}
{\large{\bf \titre }}
\bigskip \\ by \renewcommand{\thefootnote}{\star}\bigskip \\ \auteura \bigskip
\\
 \bigskip \auteurb
 \bigskip \\
\renewcommand{\thefootnote}{\dag }
\vspace{.9 in}
{\bf Abstract}
\end{center}
\abstrait
 \bigskip \\
\end{titlepage}
The Donnachie-Landshoff model of diffractive scattering$^{1-5}$ has two basic
features. Firstly, it is based on a simple physical picture,
and thus contains only a
few parameters (5 to be precise). Secondly, it works remarkably well,
accounting for all
diffractive data from 5 to 1800 GeV. In particular, it is the only model
of diffractive scattering that {\it predicted} all of the following:
the absence of a dip in $\bar p p$ elastic
collisions at the ISR, the UA4(2) and E710 values of the $\rho$ parameter, the
total and the differential elastic cross sections as observed by E710, and most
recently the total $\gamma$ cross section observed at HERA.
 These features are very different from those of leading-log
$s$ QCD, and one of the challenges of QCD is to understand
 how such a simple structure might be the
continuation of perturbative calculations to the diffractive region.

In the high-$s$ limit, the simplest behaviour for the hadronic amplitude
that one expects from Regge theory is a sum of simple poles, each pole
 corresponding
to the exchange of the particles lying on a given Regge trajectory:
\begin{equation}
{\cal A}(s,t)=\sum_i \beta_i^2(t)\ s^{\alpha_i(t)} \xi(\alpha_i(t))
 \label{hadronic}
\end{equation}
with $\xi$ the Regge signature factor.
The first (degenerate) trajectories to be included in Eq.~(\ref{hadronic})
are obviously those of
the $\rho$ and $f$ mesons.
They are the leading meson trajectories, clearly present
 when plotting the meson states in a
 $J$ vs $M^2$ plane.
When continued to negative values of $M^2$, these trajectories are responsible
for the fall-off of the total cross section at small $\sqrt{s}$.
 At higher energies, the cross sections grow, and the most natural assumption
is that another term in Eq.~(\ref{hadronic}) is responsible for that rise
although there is no observed trajectory to guide us. In fact, there is
reasonable evidence that the exchange responsible for the rise with
energy is close to a simple Regge pole, because of
 the
factorizability of the pomeron vertex$^6$: the ratio of the differential
cross section  for $p A\rightarrow XA$ to that for $pA\rightarrow pA$ does not
depend on $ A$. It is hard to imagine how a sum of poles  (or a cut)
with different couplings to $p$ and $A$ could satisfy this property.
Furthermore, Eq.~(\ref{hadronic}) leads,
through the use of the optical theorem, to total cross sections that rise like
a sum of two powers of $s$. The $pp$ and $\bar p p$ total cross sections
are indeed
well fitted, from $\sqrt{s}=5$ GeV to $1800$ GeV, to the form$^1$:
 $\sigma_{tot} = X s^{0.0808} + Y s^{-0.4525}$, with X the same for $pp$ and
 $\bar p p$, and Y changing because of the charge parity of the $\rho$
 trajectory. The two powers of $s$ seem to be universal, and are again
 encountered in all the $pX$ elastic/total cross section, for any target.
 Hence, it seems
that the pomeron is a universal object, described by a simple Regge pole with
intercept $1.08$ and charge parity $C=+1$.
 It is this result that led to the predictions
 $\sigma_{tot}^{\bar p p} (\sqrt{s}=1.8 {\rm\ TeV})\approx 73$ mb,
 $\sigma_{tot}^{\gamma p} (\sqrt{s}=200{\rm \ GeV}) \approx 160 \mu{\rm b}$ and
$\rho(\sqrt{s}=541 {\rm\ GeV})\approx \rho(\sqrt{s}=
1.8 {\rm\ TeV})\approx 0.13$.
Note that the power of $s$ describing the rise of the total cross section
is insensitive to thresholds (charm, bottom, minijets...). This fact casts
 serious doubts on models assigning the rise of the total cross section
to the opening of new inelastic channels.

If we now want to make a model for the function $\beta_P(t)$ of
Eq.~(\ref{hadronic}),
we need to address the question of the pomeron coupling.
Hadronic cross sections are observed to obey the quark counting rule, {\it
i.e.}
the cross sections are proportional to the number of valence quarks contained
in the hadron. This is tested not only in $\pi p$ vs $pp$ cross sections, but
also in cross sections involving strange quarks$^1$,
{\it e.g.} $2\sigma(\Omega^- p)
-\sigma(\Sigma^-p)$ can be predicted from $\sigma(pp)$, $\sigma(\pi p)$ and
$\sigma(Kp)$. This property indicates that the pomeron is insensitive
to the details
of the quark wavefunction, as would be the case for a coupling
to single quarks. The fact that the pomeron intercept is close to 1
then suggests that the pomeron-quark vertex has the structure $\gamma_\mu$.
This analogy
with a photon coupling leads to the conclusion that $\beta_P(t)$
 is proportional
to the
 elastic Dirac form factor $F_1(t)$, which is measured in $\gamma p$
collisions:
$\beta_P(t)=\beta_0 F_1(t)$. The fit to the total cross section fixes the value
 of $\beta_0$ to be $\beta_0\approx (0.5\ {\rm GeV)}^{-1}$.
Finally, the trajectory can
be measured via low-$t$ results for the differential elastic cross section.
ISR results from experiment R211$^7$ clearly show that
the trajectory is linear, $\alpha(t)=\alpha_0+\alpha' t$,
with $\alpha'=$(2 GeV)$^{-2}$. Note that
the analogy pomeron-photon can be pushed further$^2$ in order to describe
 single-diffraction dissociation $pp\rightarrow pX$. The idea is that
the unbroken proton
couples to the pomeron as in elastic scattering, whereas the breaking of the
other proton is described by the structure function observed in deep-inelastic
 scattering, and continued to the small-$Q^2$ region. This predicts that
the differential cross section, $d\sigma/dt dM^2$, at small $x=-t/s$, will
 behave like
$(s/M^2)^{2\alpha(t)-1} (1-M^2/s)$, where the second factor comes from
 phase-space suppression. This prediction is realized, and,
once refined$^2$ by the inclusion of the $f$ trajectory, the model
 accounts very well
 for the data at all $x$.

In order to describe the data at large $t$, or at very large $s$,
 one must worry
about $n$-pomeron exchanges. First of all, the fact that
the data are well fitted to a single pomeron pole
 indicates that
multiple exchanges are not going to play a major role in the total
cross section, at least at present energies. This is fortunate, as no one knows
how to evaluate precisely their contribution. Only their qualitative
properties are known, {\it e.g.} the fact that they will eventually dampen the
rise of the cross section and make it lower than the Froissart-Martin bound.
Instead of using an ansatz for all $n$-pomeron exchanges, Donnachie and
Landshoff have only considered two-pomeron exchange$^3$. This contribution will
turn out not to be too big, so that one may hope that further exchanges
can be neglected.
 In the context of the Donnachie-Landshoff
model, it is made of two terms: a planar one, for
 which the pomerons couple
to the same quark, and a non-planar one for which they couple to two
different quarks within the same hadron. The first process can be evaluated,
for
$t/s$ sufficiently small, as two independent scatterings in impact-parameter
space. The second one is assumed to be
 proportional to the first one, so that the
sum is a constant $\lambda$ times the second-order eikonal contribution.
In order to determine $\lambda$, one demands that the imaginary parts of
one- and two-pomeron exchanges cancel at the energy ($\sqrt{s}=31$ GeV)
and at the value of $t$ where the dip is lowest in $pp$ scattering.
This leads to $\lambda=0.43$, and to a contribution of about 10\%
 to the total cross section at S$\rm p \bar p$S/Tevatron energies. Note that
 two-pomeron exchange has a slope $\alpha'/2$ and an intercept $2\alpha_0-1$.
 The simple power of $s$ that comes from a fit
 to the total cross section must thus be considered as an {\it effective}
 power: the true one-pomeron intercept must be of the order of 0.085, and is
lowered by the two-pomeron contribution, which accounts for about $-7$ mb at
 the Tevatron. Similarly, the meson pole will include contributions
from meson-pomeron exchange$^1$.

Two-pomeron exchange is however not sufficient to reproduce the dip,
 as one still
needs to cancel the real part of the amplitude. This
is accomplished by making use of the
Landshoff mechanism$^4$,
which arises from 3-gluon exchange, each gluon being attached to a
different quark. This process is dominant at high-$t$:  when the three quarks
are scattered by the same amount, there is no price to pay for their
recombination
into a proton, whereas {\it e.g.} one-pomeron exchange would be suppressed by
$F_1(t)$. This $C=-1$, purely real contribution is calculated to be flat
in $s$ and to behave like $t^{-8}$ for large $-t$. These features are confirmed
 by data from $\sqrt{s}=27$ GeV to 53 GeV, and from $|t|\approx 3 {\rm \
GeV^2}$
to $\approx 14 {\ \rm GeV^2}$. The magnitude of this term is fixed by
 demanding a cancellation with the left-over real part of one and two-pomeron
exchange at the dip at $\sqrt{s}=31$ GeV. Its contribution gets damped at small
$t$ by a regularization of the gluon propagator.
As the cancellation in the dip region is largely due to the 3-gluon exchange
 term, and as that term is odd, the model predicted that there would not be
a dip in $\bar p p$ scattering. That has now been verified at the ISR$^8$.
 Note that this is
the {\it only} evidence for an odderon contribution in hadronic scattering.

 Finally, let us mention that this model also makes definite predictions
concerning the photon and proton structure functions$^5$.
Indeed, in $\gamma^* p$
scattering, the same hadronic amplitude gets probed, the only change being
that it gets attached to an upper photon line. The energy flowing through it
is given by $s=-(k^2+k_\perp^2)/ x -k^2+(1-x) m_p^2$, where $k^2$ is the
off-shellness of the quark attached to the photon, $k_\perp$ its transverse
momentum, and $ x$ the fraction of the proton momentum that it carries. Hence,
at very small $x$, the effective $s$ is very large, the process is
dominated by pomeron exchange, and the structure functions are predicted
to behave like $1/x^{1.08}$.
One might point out that models that neglect
the off-shellness of the quark and/or its transverse momentum, are bound to
miss this behaviour. A recent parametrization of all the pre-HERA data
consistent with the above ideas is now available$^5$.

We have thus seen that the Donnachie-Landshoff model of diffractive scattering
reproduces all data from $\sqrt{s}=5$ GeV to present energies, for
the scattering of protons onto $p, \bar p, \pi, K,\gamma,d$. It also provides
 a successful picture of diffractive
dissociation, and,  as explained elsewhere$^9$, of the pomeron structure
 function.
 Let us nevertheless mention that two recent measurements are in
conflict with the model$^{10}$: H1 observes a much steeper rise of the
structure
 functions at small-$x$, and CDF measures a cross section which is
larger than the E710 number predicted by the model. If confirmed, these data
would indicate that something new is happening, maybe the long-awaited
 perturbative contribution. One would then have to
decide how to subtract the non-perturbative contributions
 from the data, in order to make meaningful comparisons with
theory.
This question leads us naturally to the second part of this paper, which deals
with the possibility of obtaining a description of non-perturbative scattering
 from QCD.

It is now well-known that the leading-log $s$ resummation$^{11}$
cannot account for diffractive scattering. The first problem is that
the amplitude is not factorizable: this is in fact a consequence of the
infrared finiteness of the answer. Quark-quark scattering via gluon
exchange diverges for massless gluons. Nevertheless, hadron-hadron scattering
is infrared finite, as the colour of the hadron gets averaged for very long
wavelength gluons. Hence there is a contribution that comes from the diagrams
where gluons are exchanged between different quarks in the hadron. These
diagrams feel the hadronic wavefunction, and hence their contribution depends
on
the target. A second problem is that the rise of the cross section predicted by
perturbative QCD is entirely inappropriate to describe data, as its leading
contribution to the hadronic amplitude goes like
$s^{1+2.65\alpha_s}$. Finally, we have seen that hadronic data contain
a scale of the order of  $ 1/\sqrt{\alpha'}\approx 2$ GeV, which comes in the
description of the $t$-dependence of the hadronic amplitude. No such scale is
 present in perturbative QCD, hence the differential elastic cross section
has the wrong shape:
its logarithmic slope at $t=0$ is infinite and its curvature is too big.

 One is thus
led to the semantic distinction of a ``soft pomeron'', which describes the
data at low momentum transfers, and a ``hard pomeron'', which is supposed to
arise once $\alpha_s$ is small enough and $s$ is big enough. No one knows how
these objects would combine, and quite a few alternatives have been proposed in
 the literature$^{12}$. We want here to address the question of the matching of
a non-perturbative pomeron with a perturbative one at the level of quarks
and gluons, {\it i.e.} we want to consider what ingredients would transform
the perturbative calculation into a viable model of non-perturbative
 exchanges$^{13}$.

The first question concerns the factorizability of the amplitude.
The $n$-gluon
contribution to the hadronic amplitude ${\cal A}_h$ takes the form:
\begin{equation}
{\cal A}_{h}(s,t) =\sum_n\int\alpha_S^n \left[\prod_{i=1}^n dk_\perp^i \right]
{\cal A}_{qq}(k_\perp^i)\
 \left[{\cal E}_1^2(t)-{\cal E}_d(k_\perp^i)\right]
\label{factorize}
\end{equation}
where ${\cal A}_{qq}$ is the quark-quark scattering amplitude,
${\cal E}_1^2(t)$ is
the form factor associated with the exchange when the gluons are attached to
the
same quark line $-$ one can show that ${\cal E}_1$ is equal to the Dirac
form factor $F_1\ -$, and ${\cal E}_d$ comes from the other diagrams,
where the gluons are attached to different quarks. It is the latter that
guarantees the infrared finiteness of the answer, and that
 produces violations
of factorizability. Hence this term must be suppressed by non-perturbative
 effects.
It contains explicitly the hadron wavefunction, and thus a scale $R$,
the hadronic radius. Landshoff and Nachtmann$^{14}$ have observed that an
infrared suppression of the gluon propagator would indeed generate a
suppression. If the gluon propagator is smoother than a pole in the infrared
region, then on dimensional grounds, it must contain a scale $\mu_0$:
$D(q^2)={(1/\mu_0^2)}\ {\cal D}({q^2/\mu_0^2})$. If the scale $\mu_0$ is big
enough, then the propagator does not change much while the form factor
${\cal E}_d$ drops sharply. Hence, for $\mu_0>>1/R$, one gets a suppression
proportional to $1/(\mu_0 R)^2$.

This idea that the gluon propagator should be smoother than the
perturbative one has received some theoretical confirmation from lattice
gauge theory$^{15}$, from studies of the Dyson-Schwinger equations$^{16}$
 and from considerations related to the Gribov Horizon$^{17}$. Changing the
gluon propagator amounts to the inclusion of a subclass of diagrams
(the gluon self-energy) which are  resummed via non-perturbative
methods. All these propagators give rise to a factorizable
amplitude for $\mu_0$ big enough.

The next question is whether the scale of the amplitude is correct. It is
already known that purely perturbative 2-gluon exchange$^{18}$ gives reasonable
numbers for the total cross section, but leads to the wrong shape for the
differential elastic cross section.
 Although the various non-perturbative propagators
 are far from agreeing, for a scale $\mu_0$
of the order of 0.5 to 1 GeV, each can give a good starting value
for the total cross section, of the order of 20 mb.
The logarithmic slope of the elastic cross section also gets cured by the
introduction of $\mu_0$, and numbers of the order of 10 GeV$^{-2}$ can
be obtained.
At this order, as has now been observed by many authors$^{19}$,
the inclusion of modified propagators in the calculation provides
appreciable improvements and the improved order $\alpha_s^2$ constitutes
a good starting point for an expansion
in $\log s$. Let us now see to which extent these improvements carry over to
higher orders$^{13}$.

The complete order $\alpha_s^3$ leads to a hadronic amplitude
that has the following form:
\beq
\amp(s,t)= \amp_2\left\{ i\left[1+\log s \left(\epsilon_0+\alpha
't+O(t^2)\right)\right]+f_{odd}(t)\right\}
\label{structure}
\eeq
It is of course tempting to see in this a first-order Taylor expansion in
$\log s$ of a pomeron pole, plus a zeroth-order term from an odderon pole.
 As BFKL have shown$^{11}$, life is not so simple,
and higher-order terms spoil the analogy. Hence, in the following,
the terms ``pomeron intercept'' or ``slope'' must not be taken literally.
At this order,
the normalization of the cross section, $\amp_2$,
comes from two-gluon exchange; in the Feynman gauge, the coefficient of $\log
s$, $\epsilon_0$, is the ratio of two to three gluon exchange.
 The odd contribution $f_{odd}(t)$
 comes entirely from 3-gluon exchange.
 Finally, the $\alpha '$ contribution comes from the diagrams involving the
3- and 4-gluon vertices and from the Taylor expansion of 3-gluon exchange.
Note that one can show that, at least for 3-gluon exchange diagrams,
the replacement of the perturbative gluon propagator by a non-perturbative
counterpart does not violate gauge invariance and can be justified
theoretically$^{13}$, provided that the propagators have a Hilbert transform.

Let us first consider the odd term $f_{odd}$. It is purely real,
 and proportional to $s$. As expected, it contains a
contribution from the Landshoff mechanism. The structure of its form factor
clearly shows that the Landshoff term is dominant at high-$t$ whereas it
contributes  about 18\% of the two-gluon exchange amplitude at $t=0$. It
is present only in baryon-baryon scattering, and not in meson-baryon
scattering.

The two main problems of the perturbative expansion are already manifest at
this
order when one uses perturbative propagators: the pomeron intercept is
$1+1.85 \alpha_s$, and the slope turns out to be logarithmically divergent,
although $\alpha' t\rightarrow 0$ when $t\rightarrow 0$.
Non-perturbative gluon propagators bring the calculation in closer
 agreement with the data, but
their effect is not sufficient:  any
substantial reduction of the order $\alpha_s^3$ result also brings the
 normalization of the cross section down. For values of $\mu_0$ favoured
 by our previous
discussion on two-gluon exchange,  we get values of the order of 2
for the intercept, and it seems impossible to get acceptable numbers
both for two and three-gluon exchange. Similarly,
 the pomeron slope becomes finite once the infrared
region is smoothed out. Values compatible with 0.25 GeV can again be achieved,
but again for values of $\mu_0$ or $\alpha_s$ that would suppress
two-gluon exchange. Hence, one cannot
accommodate a sizeable two-gluon exchange amplitude together
with a slowly rising third-order amplitude by simply modifying the gluon
propagator according to the results of Ref.~16.

Nevertheless, the inclusion of non-perturbative propagators inside the
perturbative calculation has a rather large effect: qualitatively, it replaces
the infinities of the perturbative answer ($B(0),\ \alpha'$) by finite
numbers, and gives rise to a factorizing amplitude. Quantitatively,
it reduces $\epsilon_0$ by a factor of the order of 2,
and gives good starting values for the order $\alpha_S^2$.

 It is worth pointing out that
the factorizability of the pomeron vertex, the smallness of the intercept and
the pomeron slope point to rather large values
of the scale $\mu_0$. The problem would then be to find a mechanism to increase
the two-gluon exchange term. It is conceivable that contributions from the
color field would increase all orders by a common factor (as they would act
at the form-factor level), and one would then need to use a larger value
of $\mu_0$, hence suppressing the higher orders more and more
(as the number of propagators grows with the order).

It is also possible
that the standard techniques perturbation theory (such as cutting rules) become
modified in the non-perturbative regime. This could totally modify the
estimate of higher orders. Indeed, one finds that the diagrams giving rise to
 $\epsilon_0$ are made of two contributions: a cut through the quarks, and
 another cut
 through the quarks and one gluon. Were the latter to be diminished by a factor
 2, $\epsilon_0$ would become identically zero.

These problems are presently under further investigation.

\vglue 0.6cm
{\bf \noindent Acknowledgements \hfil}
\vglue 0.4cm I thank Peter Landshoff for his comments and suggestions,
 and Carmen A. Pont for a careful proofreading.
 This work was supported in part by NSERC (Canada) and les fonds FCAR
(Qu\'ebec).
%\end{enumerate}

\end{document}